\documentclass[10pt,letterpaper]{article}
\usepackage{hyperref}
\usepackage{amsmath}
\usepackage{subfigure}
\usepackage{epsfig}

\begin{document}

%\twocolumn[
\title{\bf Illusion Media: Generating Virtual Objects Using Realizable Metamaterials}

\author{Wei Xiang Jiang, Hui Feng Ma, Qiang Cheng and Tie Jun Cui\footnote{tjcui@seu.edu.cn}\\
{\small\it State Key Laboratory of Millimeter Waves and Institute of Target Characteristics and Identification} \\
{\small\it Department of Radio Engineering, Southeast University,
Nanjing 210096, P. R. China.}}

\date{}
\maketitle

\begin{abstract}

We propose a class of optical transformation media, illusion media,
which render the enclosed object invisible and generate one or more
virtual objects as desired. We apply the proposed media to design a
microwave device, which transforms an actual object into two virtual
objects. Such an illusion device exhibits unusual electromagnetic
behavior as verified by full-wave simulations. Different from the
published illusion devices which are composed of left-handed
materials with simultaneously negative permittivity and
permeability, the proposed illusion media have finite and positive
permittivity and permeability. Hence the designed device could be
realizable using artificial metamaterials.

\hskip 1.0mm


\end{abstract}
%]

\newpage

Since Pendry \emph{et} al. and Leonhardt proposed an interesting
idea to design invisibility cloaks [1,2], more and more attention
has been paid to the cloaks and other optical transformation devices
[3-16]. Recently, Lai \emph{et} al. have presented a concept of
illusion optics: making an object with arbitrary shape and material
properties appear exactly like another object of some other shape
and material makeup [13]. Invisibility cloak can be regarded as
generating an illusion of free space [1-5]. Using the transformation
optics, they designed an illusion device consisting of two distinct
pieces of metamaterials, which are called \emph{complementary
medium} and \emph{restoring medium}. However, the complementary
medium was obtained from the transformation optics of folded
geometry. Hence it is composed of left-handed materials with
simultaneously negative permittivity and permeability. As a result,
the proposed illusion device is extremely demanding of material
parameters, and the applications are possible to remain in the realm
of theory [18].

In this letter, we present a class of optical transformation media
which can make the enclosed actual object invisible and generate one
or more virtual objects at the same time. An arbitrary object
enclosed by such a medium layer appears to be one or more other
objects of arbitrary shapes and material makeup. As applications, we
design an illusion device, which transforms an actual object into
two virtual objects. Such a device exhibits unusual electromagnetic
(EM) behaviors as verified by full-wave simulations. Here, we try to
make the illusion media be fairly realizable. Unlike the published
illusion devices which are composed of left-handed materials [13],
all permittivity and permeability components of the proposed
illusion media are finite and positive. Hence the presented approach
makes it possible to realize the illusion media using artificial
metamaterials. The principle behind such illusion media is not only
bending EM waves around the actual object, but also generating one
or more virtual objects within the virtual boundary.

An intuitive schematic to illustrate the proposed method is
illustrated in Fig. 1. A golden apple (the actual object) is
enclosed with an illusion medium layer, as shown in Fig. 1(a). Such
a layer of illusion media makes any detector outside the virtual
boundary (the dashed curves) perceive the scattering fields of two
green apples (the virtual objects, shown in Fig. 1(b)) instead of
one golden apple. In the other word, the illusion medium layer makes
the EM fields outside the virtual boundary in both the physical and
virtual spaces exactly the same, regardless the direction of the
incident waves. The illusion medium layer has two functions,
concealing the optical signature of the golden apple and generating
the image of two green apples. The coordinate transformation of the
illusion medium layer is similar to that of an invisibility cloak,
but the illusion space contains some virtual objects with given EM
parameters instead of purely free space. The permittivity and
permeability tensors of the illusion media are calculated by
\begin{eqnarray}
\overline{\varepsilon'}=\mathrm{\Lambda}\overline{\varepsilon}
\mathrm{\Lambda}^{T}/\det(\mathrm{\Lambda}),~~~~
\overline{\mu'}=\mathrm{\Lambda}\overline{\mu}
\mathrm{\Lambda}^{T}/\det(\mathrm{\Lambda}),
\end{eqnarray}
in which $(\overline{\varepsilon}$, $\overline{\mu})$ and
$(\overline{\varepsilon'}$, $\overline{\mu'})$ are the constitutive
tensors in virtual space (illusion space) and the physical medium
layer, respectively, and $\mathrm{\Lambda}$ is the Jacobian
transformation matrix with components
$\mathrm{\Lambda}_{ij}=\partial x_{i}/\partial x_{j}$, corresponding
to the mapping from the illusion space to the illusion medium layer.

Similar to invisibility cloaks, the EM fields in the illusion medium
layer can be found from the transformation optics [1,2] as
$\mathbf{E}'=(\mathrm{\Lambda}^T)^{-1}\mathbf{E}$ and
$\mathbf{H}'=(\mathrm{\Lambda}^T)^{-1}\mathbf{H}$, where
$\mathbf{E}$ and $\mathbf{H}$ are electric and magnetic fields in
the virtual space, respectively. Because the boundary $s_2$ is
mapped to itself during the transformation $\mathrm{\Lambda}$, we
have $\mathbf{E}'_t=\mathbf{E}_t$ and $\mathbf{H}'_t=\mathbf{H}_t$,
where the subscript $t$ indicates transverse components along the
surface $s_2$. In another word, the tangential components of the EM
fields on the whole virtual boundary ($s_2$) are exactly the same in
the physical and virtual spaces. Hence, by the uniqueness theorem,
the EM fields outside are also exactly the same. Any detector
outside the illusion medium layer will perceive EM waves as if they
were scattered from the illusion object (two green apples and
nothing else), and thus an illusion is generated. In the design of
invisibility cloaks, the illusion space is only free space, hence
there were almost no scattered fields from a perfect cloak. For the
illusion media, however, the illusion space may contain many
illusion objects of our choice, thus the scattered fields are the
same as those from the virtual objects.

In order to demonstrate the design of illusion medium layer, we take
an illusion device as example and make full-wave simulations based
on the finite element method. We consider an illusion medium layer
which transforms a metallic square cylinder to two dielectric square
cylinders. In the following simulations, we consider the case when a
transverse-electric (TE) polarized plane wave is incident upon an
illusion medium layer, hence there exists only $z$ component of
electric field. The case is very similar when a transverse-magnetic
(TM) wave is incident.

Figure 2 illustrates the numerical results of scattered electric
fields for an illusion medium layer which transforms one metallic
square cylinder to two smaller dielectric square cylinders. In order
to illustrate the scattering character of objects better, we have
plotted the scattered fields instead of total fields here and next.
The plane waves are incident horizontally from the left to the right
at 12 GHz. Figs. 2(a) and 2(c) show the scattered patterns of the
single metallic square cylinder without the illusion medium layer
and two dielectric square cylinders with $\varepsilon_r=3$ and
$\mu_r=2$, respectively. When enclosed by the illusion medium layer,
the scattered pattern from the single metallic square cylinder will
be changed as if there were two smaller dielectric square cylinders.
This can be clearly observed by comparing the scattering-field
pattern of the metallic cylinder coated by the illusion medium layer
shown in Fig. 2(b) with that of two dielectric cylinders shown in
Fig. 2(c). The field patterns are exactly the same outside the
virtual boundary. Inside the virtual boundary, the field patterns in
Figs. 2(b) and 2(c) are different.

Unlike the earlier-proposed illusion devices which consist of two
distinct pieces of metamaterials - the complementary medium with
double negative constitutive parameters and the restoring medium
[13], the illusion medium layer in Fig. 2(b) is only composed of one
layer of inhomogeneous and anisotropic medium. To design such an
illusion medium layer, we first consider a three-dimensional (3D)
one obtained by rotating the mirror-symmetric square around the
\emph{x}-axis. We construct the transformation for each side in the
Cartesian coordinate system, which can be expressed as
\begin{eqnarray}
\frac{x'}{x}=\frac{y'}{y}=\frac{z'}{z}=k(x,y),
\end{eqnarray}
where $k(x,y)=k_0-c/(ax+by)$ and $k_0$ is a compression factor,
i.e., the ratio of the sizes of actual and virtual spaces. We have
assumed that the corresponding side of polygonal metallic cylinder
be expressed as $ax+by+c=0$. In this example, the side length of
each virtual object is 0.02 m and $k_0=0.5$. Based on the above
transformation, the permittivity tensor of the illusion medium layer
can be calculated. We can obtain two-dimensional (2D) parameters by
setting $z=0$,
\begin{eqnarray}
&&\varepsilon'_{xx}=\varepsilon_r(\lambda_{11}^2+\lambda_{12}^2)/\lambda_{33},\hskip 20mm\\
&&\varepsilon'_{yy}=\varepsilon_r(\lambda_{21}^2+\lambda_{22}^2)/\lambda_{33},\\
&&\varepsilon'_{xy}=\varepsilon_r(\lambda_{11}\lambda_{21}+\lambda_{12}\lambda_{22})/\lambda_{33}=\varepsilon'_{yx},\\
&&\varepsilon'_{zz}=\varepsilon_r/k_0,~~\varepsilon'_{xz}=\varepsilon'_{yz}=\varepsilon'_{zx}=\varepsilon'_{zy}=0,
\end{eqnarray}
where $\lambda_{11}=k_0-bcy/(ax+by)^2$, $\lambda_{12}=
bcx/(ax+by)^2$, $\lambda_{21}= acy/(ax+by)^2$,
$\lambda_{22}=k_0-acx/(ax+by)^2$, and $\lambda_{33}
=k_0\big(k_0-c/(ax+by)\big)^2$. Obviously, the component
$\varepsilon'_{zz}$ is constant. The magnetic permeability can be
expressed in a similar way, and we omit the expressions here. We
remark that a 3D transformation have been constructed in equation
(2) instead of a 2D transformation to eliminate the singularity of
the EM parameters [11,17].

We have assumed that the illusion objects are isotropic:
$\overline{\varepsilon}=\varepsilon_{r}\overline{I}$ and
$\overline{\mu}=\mu_{r}\overline{I}$. For two trapezia regions in
Fig. 2(b), $\varepsilon_r=3$ and $\mu_r=2$, which generate the
illusion of two square dielectric cylinders; for other regions in
the illusion medium layer, $\varepsilon_r=1$ and $\mu_r=1$, which
generate the illusion of free space. Due to the non-conformality of
the transformation, some non-diagonal components of the constitutive
tensors are non-zero. However, the real fabrication requires the
material parameters $\overline{\varepsilon'}$ and $\overline{\mu'}$
to be denoted in diagonal tensors. The symmetry of the tensors
$\overline{\varepsilon'}$ and $\overline{\mu'}$ indicates that there
always exists a rotation transformation which maps a symmetric
tensor into a diagonal one. Therefore, the material parameters with
the eigenbasis can be expressed as
\begin{eqnarray}
&&\varepsilon_{1}=
\frac{\varepsilon_{xx}^{'}+\varepsilon_{yy}^{'}-\sqrt{\varepsilon_{xx}^{'2}-2\varepsilon_{xx}^{'}\varepsilon_{yy}^{'}+\varepsilon_{yy}^{'2}+4\varepsilon_{xy}^{'}}}{2}, \hskip 20mm\\
&&\varepsilon_{2}=\frac{\varepsilon_{xx}^{'}+\varepsilon_{yy}^{'}+\sqrt{\varepsilon_{xx}^{'2}-2\varepsilon_{xx}^{'}\varepsilon_{yy}^{'}+\varepsilon_{yy}^{'2}+4\varepsilon_{xy}^{'}}}{2},\\
&&\varepsilon_{z}=\varepsilon_{zz}^{'},
\end{eqnarray}
and all non-diagonal components are zero. The magnetic permeability
$\mu_{1}$, $\mu_{2}$, and $\mu_{z}$ have similar expressions, and
are omitted here. We remark that the relation between the virtual
and physical coordinates can be established as
$x/x'=y/y'=z/z'=\big(1+c/(ax'+by')\big)/k_0$.

For TE-wave incidence, only $\mu_1$, $\mu_2$ and $\varepsilon_z$ are
of interest and must satisfy the request of Eqs. (3)-(9). The
distributions of $\mu_1$ and $\mu_2$ are illustrated in Figs.
3(a)-(b), from which we clearly see that all values are finite
without any singularity. It is important to note that the
transformation in this illusion device is a compressing mapping,
instead of the folding of geometry [10,13,16]. Hence any parameters
of the illusion medium layer is not negative. This kind of
metamaterials have been extensively studied and fabricated in the
experiment of free-space cloaks at microwave frequency [3]. Similar
to the experiment of the free-space cloak, such an illusion medium
layer is restricted to a narrow frequency band because some
parameter components go to zero near the inner boundary [3,15]. The
illusion medium layer renders the enclosed metallic square cylinder
invisible and projects the illusion of two dielectric square
cylinders as shown in Fig. 2(b).

In real applications, artificial metamaterial structures are always
lossy. Hence it is interesting to investigate the lossy effect of
the medium layer on the illusion property. When we add a loss
tangent of 0.01 in both permittivity and permeability of the
illusion medium layer, the scattered electric-field distributions
(not shown) are very similar to that of two dielectric square
cylinders. Hence, the lossy illusion medium layer is still
effective.

The presented illusion medium layer could be realized by designing
proper metamaterial structures. To achieve the permeability
component $\mu_1$, we can make use of the split-ring resonators
(SRR) [3,11]. One kind of SRR unit, the C structure, has been used
for the design, as shown in Fig. 4(a). From the retrieved
permeability illustrated in the same figure, we could obtain the
permeability component ranging from 0 to 0.7 by adjusting the SRR's
size, which satisfies the requirement to $\mu_1$ shown in Fig. 3(a).
Another kind of SRR structure, which can be used to achieve the
other permeability component $\mu_2$, is demonstrated in Fig. 4(b).
Clearly, the retrieved permeability varies from 2 to 5.2 by changing
the SRR's size, satisfying the requirement to $\mu_2$ shown in Fig.
3(b). The component $\epsilon_z=2$ or 6 can be realized using the
non-resonant structure - I shape [5]. All the imaginary parts of
electromagnetic parameters for the metamaterial units at the
non-resonant frequencies are small enough to be neglected in the
design of illusion medium layer. We remark that it is indeed a
complicated work to optimize the overall design to make a compact
layout of the illusion medium layer, which will be considered in the
further work.

In summary, we have presented a class of optical transformation
media, illusion media, which can create one or more virtual objects
by using metamaterials. To eliminate the singularity of medium
parameters for the illusion device, we construct the transformation
in 3D. Hence, in such illusion media, all components of constitutive
parameters in the principle coordinate system are finite and
positive and they could be realizable using artificial structures.

This work was supported in part by the National Science Foundation
of China under Grant Nos. 60990320, 60990324, 60871016, 60921063 and
60901011, in part by the Natural Science Foundation of Jiangsu
Province under Grant No. BK2008031, and in part by the 111 Project
under Grant No. 111-2-05. WXJ acknowledges the support from the
Graduate Innovation Program of Jiangsu Province under No.
CX08B\_074Z and the Scientific Research Foundation of Graduate
School of Southeast University under No. YBJJ0816.

%\newpage

%\newpage

\section*{{{\bf List of Figure Captions}}}

\noindent \textbf{Fig. 1:} {(color online) A simple scheme of an
illusion medium layer that transforms the image of an object (a
golden apple) into that of the illusion (two green apples). (a) The
golden apple (the actual object) enclosed with the illusion medium
layer in the physical space. (b) Two green apples (the illusion) in
the virtual space.}

\vskip 5mm

\noindent \textbf{Fig. 2:} {(color online) The scattered
electric-field distributions in the computational domain for (a) a
bare metallic square cylinder; (b) a metallic square cylinder with
the illusion medium layer; and (c) two dielectric square cylinders
when the plane waves are incident horizontally.}

\vskip 5mm

\noindent \textbf{Fig. 3:} {(color online) The parameter
distributions of the illusion medium layer, (a) $\mu_1$, (b)
$\mu_2$.}

\vskip 5mm

\noindent \textbf{Fig. 4:} {(color online)  The effective EM
parameters versus to the frequency for the metamaterial structures.
(a) C structure for $\mu_1$; (b) SRR structure for $\mu_2$.}

%\end{document}

\vskip 30mm

%\newpage

\begin{figure}[h,t,b]
\centerline{\includegraphics[width=10cm]{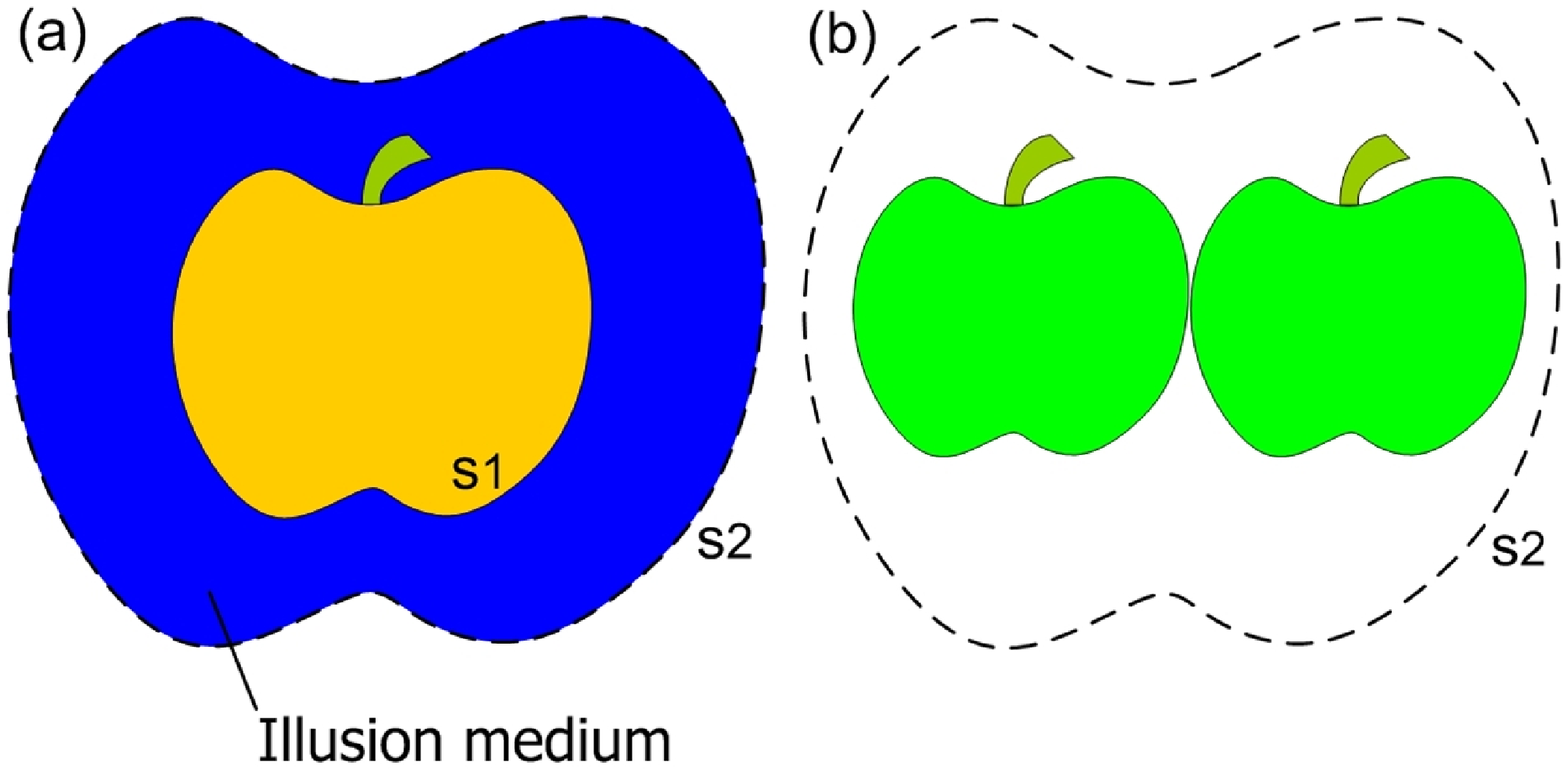}}
 \caption{\small
}
\end{figure}

%\newpage

\begin{figure}[h,t,b]
\centerline{\includegraphics[width=10cm]{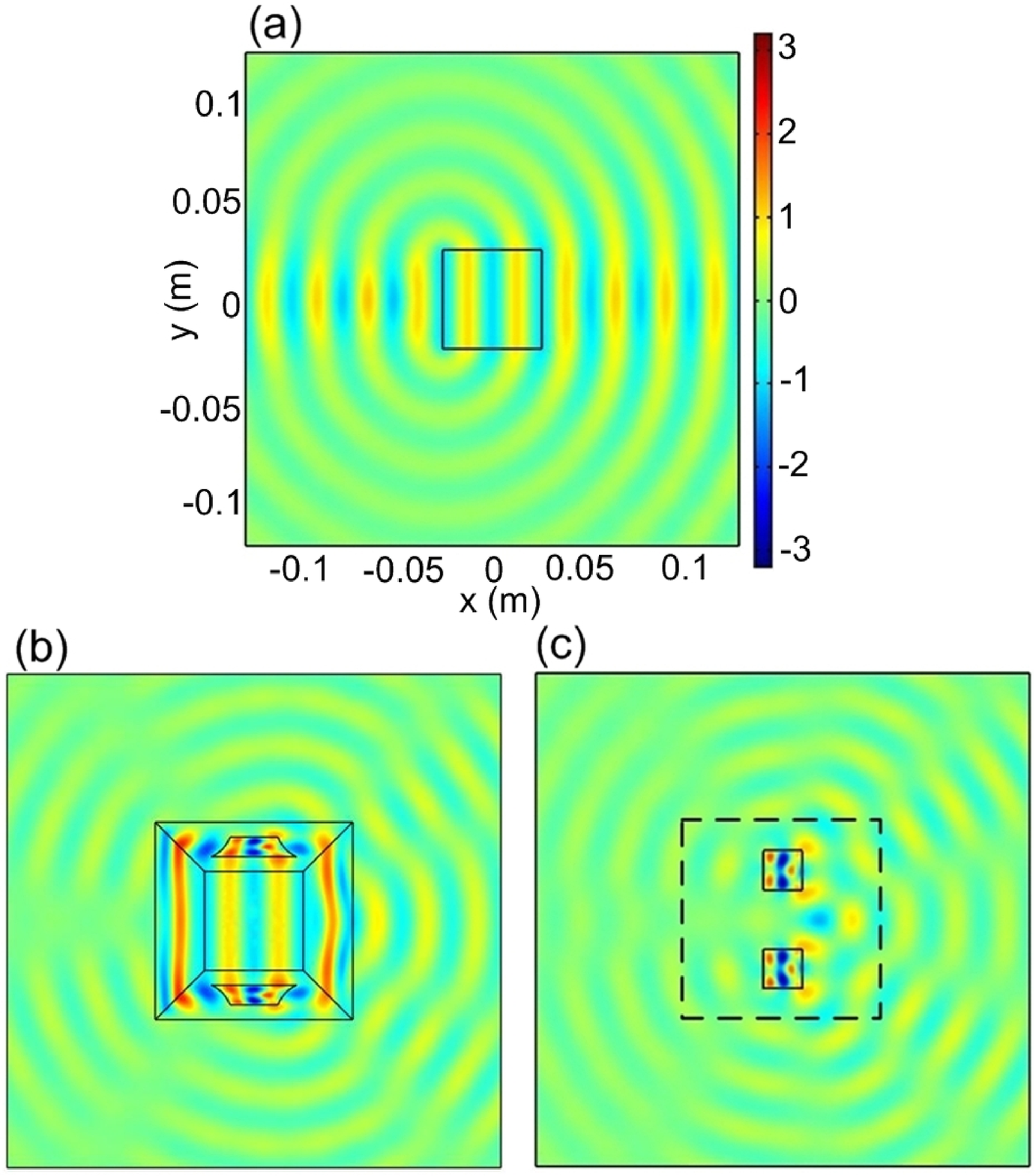}}
 \caption{\small
}
\end{figure}

\newpage

\begin{figure}[h,t,b]
\centerline{\includegraphics[width=12cm]{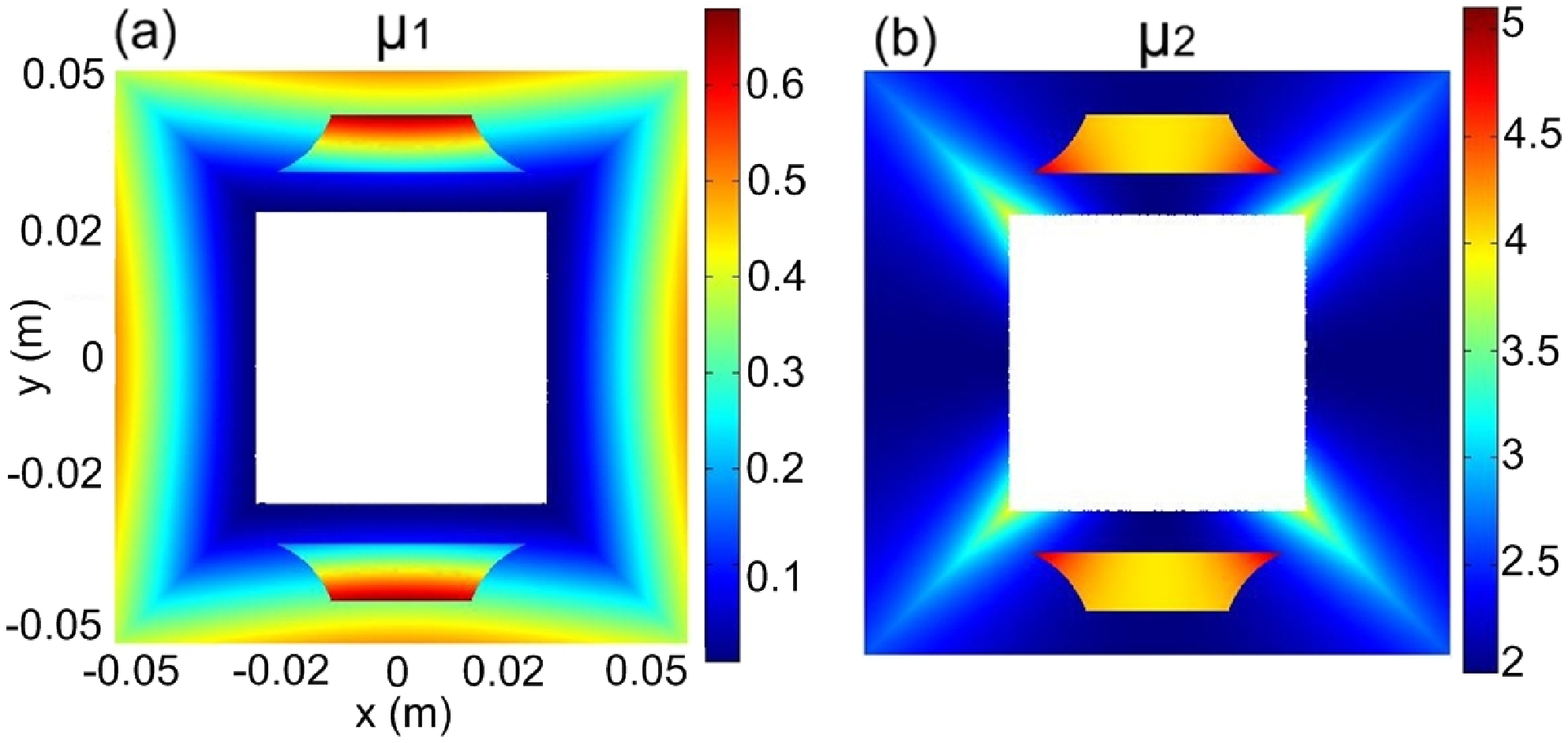}}
 \caption{\small
}
\end{figure}

%\newpage

\begin{figure}[h,t,b]
\centerline{\includegraphics[width=12cm]{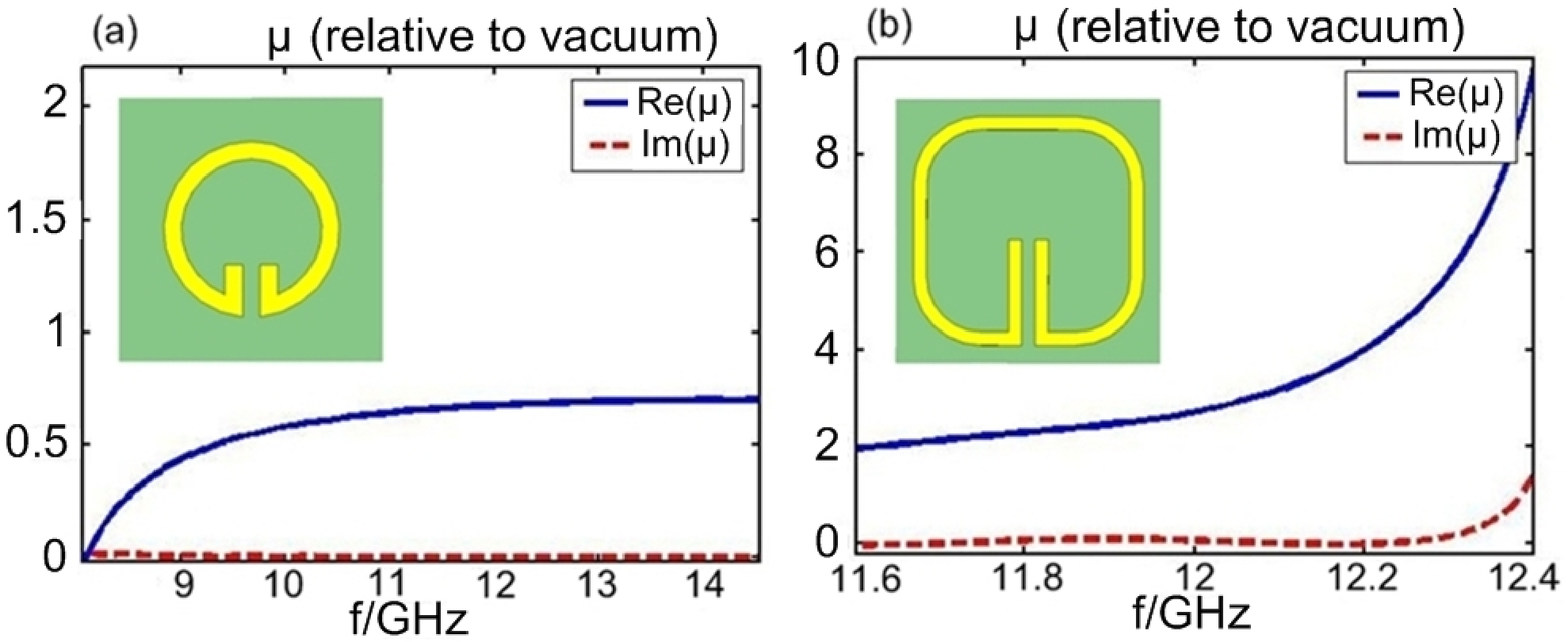}}
 \caption{\small
}
\end{figure}

\end{document}